# Policy Enforcement with Proactive Libraries

Oliviero Riganelli, Daniela Micucci, and Leonardo Mariani
Department of Informatics, Systems and Communication
University of Milano Bicocca
Viale Sarca 336, IT-20126 Milan, Italy
Email: {riganelli,micucci,mariani}@disco.unimib.it

*Abstract*—Software libraries implement APIs that deliver reusable functionalities. To correctly use these functionalities, software applications must satisfy certain correctness policies, for instance policies about the order some API methods can be invoked and about the values that can be used for the parameters. If these policies are violated, applications may produce misbehaviors and failures at runtime. Although this problem is general, applications that incorrectly use API methods are more frequent in certain contexts. For instance, Android provides a rich and rapidly evolving set of APIs that might be used incorrectly by app developers who often implement and publish faulty apps in the marketplaces.

To mitigate this problem, we introduce the novel notion of *proactive library*, which augments classic libraries with the capability of proactively detecting and healing misuses at runtime. Proactive libraries blend libraries with multiple proactive modules that collect data, check the correctness policies of the libraries, and heal executions as soon as the violation of a correctness policy is detected. The proactive modules can be activated or deactivated at runtime by the users and can be implemented without requiring any change to the original library and any knowledge about the applications that may use the library.

We evaluated proactive libraries in the context of the Android ecosystem. Results show that proactive libraries can automatically overcome several problems related to bad resource usage at the cost of a small overhead.

*Keywords*-proactive library; self-healing; Android; resource usage; API; policy enforcement

## I. INTRODUCTION

Software development for the mobile market has resulted in multiple ecosystems regularly accessed by a huge number of mobile device users. An ecosystem is typically composed of a marketplace (e.g., the Android's Google Play, the Apple's App Store, and the Microsoft's Windows Phone App Store), application providers (i.e., who implement and distribute apps through the marketplace), and customers (i.e., who download, install, and use apps available in the marketplace) [1].

The marketplace is open to everyone who likes to contribute by making apps available for download. However, the openness of such ecosystems raises reliability issues. In fact, the apps that are available on the marketplaces are not necessarily implemented by professional developers and might be unsafe and unstable. For instance, developers may accidentally implement and release faulty apps, which might cause issues in the devices where they are installed. To mitigate these problems, the marketplaces might be equipped with compensation mechanisms. For example, users can rate and review apps, so that unsafe and unreliable apps become less and less popular and fewer people try to install and use them [2]. Ecosystems may also implement static analysis routines to reject largely inadequate apps [3].

Although these mechanisms can be helpful to prevent the proliferation of undesired apps, customers regularly report issues with the apps downloaded from the marketplaces [4], [5], [6], [7], [8]. A relevant portion of the problems experienced by the users is related to the way apps interact with the resources available in a device (e.g., the camera, the microphone, and the wifi antenna). For instance, many apps fail to properly acquire and release resources, causing efficiency and energy problems [4], [5], [8]. In some cases, inaccurate interactions with the resources may even cause problems across apps. For instance, an app that does not release the camera every time its execution is suspended may prevent the other apps from acquiring the camera [4], [8].

Interestingly the incorrect interaction between an app and a resource *can be often recognized by looking at the lifecycle of the activities[1] that compose the app and the calls to the API that controls the access to that resource*. For instance, if the user moves to background an app that is using the microphone to record some audio, the app must immediately release the microphone, otherwise the other apps might be unable to interact with the microphone because it would be occupied by the inactive app, and the device may unnecessarily keep the microphone active consuming extra battery. In the Android ecosystem, this means that once an app has invoked the method `startRecording()` on the `AudioRecord`, it must release the microphone by invoking the method `release()` every time a call to the `onStop()` callback method is generated by the Android framework (`onStop()` is a callback method that apps implement to define the operations that must be performed before an activity becomes invisible to the user). Unfortunately, it is not always the case that apps are disciplinely implemented releasing and acquiring resources coherently with the lifecycle of the activities [9], [4], [8].

To prevent users from experiencing annoying problems caused by apps that inaccurately interact with APIs, we introduce a novel concept of library that we called *proactive library*. A proactive library is a library composed of two main parts: the reactive library, which is a regular implementation of an API, and a set of proactive modules, which decorate the library with the ability to enforce some *correctness policies* at runtime, de facto augmenting the system with self-healing capabilities.

---
[1]Android apps are composed of multiple components called activities. https://developer.android.com/guide/components/activities

The proactive modules operate jointly with their library reacting to the invocation of certain API and callback methods, checking that the apps are using the API appropriately, and automatically fixing the execution and the status of the system, if necessary. For example, when the `onStop()` callback is generated, the proactive module of the library for audio recording is triggered for checking if the microphone has been released. If not, the proactive module may force the release of the microphone, and automatically reassign it again to the activity once the activity becomes visible to the user again.

Since a regular library should not be changed to be turned into a proactive library, the proactive modules can be easily added to existing libraries, without the need to design a library as a proactive library from the beginning. This eases the incremental adoption of the technology. Moreover, any party may contribute to the definition of proactive modules for a given library and not only the developers of the library.

From an end-user perspective, the replacement of regular libraries with proactive libraries gives to users the opportunity to make their systems more stable, at the cost of a small overhead that might be introduced if many correctness policies are simultaneously violated. Moreover, the proactive modules can be turned on or off by users who can decide how to execute their applications case by case. This decision depends on the level of trusting on the installed apps and past experience. For instance, if users intend to use an app developed by an untrusted source or an app that caused problems in the past, they may activate the proactive modules before running the app, otherwise they may keep the proactive modules turned off and use the app normally.

To support development activities, we defined a model-based approach for the definition and generation of proactive libraries from correctness policies. We experienced our approach on Android, and focused on policies related to resource usage. Although the concept of proactive library and correctness policy are general and not specific to the Android ecosystem, we think this solution is particularly relevant in the context of mobile systems where many non-trivial and rapidly evolving APIs for resource management are available. Our empirical evaluation based on policies for six different Android resources shows that proactive libraries can be a useful design approach to prevent problems at runtime.

In a nutshell this paper provides the following contributions:
- It introduces the notion of *proactive library*, which is a library decorated with proactive modules that can provide self-healing abilities for certain correctness policies,
- It presents a *formalism* for the specification of the policy enforcement strategies and provides guidelines about the *generation of the Java proactive modules* from their formal definitions,
- It provides *empirical evidence* about the effectiveness and efficiencies of proactive libraries.

The paper is organized as follows. Section II introduces a running example that is used throughout the paper to exemplify the approach. Section III presents the concept of proactive library and the process for implementing it. Section IV describes how to formally define the correctness policies that can be enforced with proactive libraries. Section V presents how to obtain the proactive module from the specification of the correctness policies that must be enforced. Section VI presents our empirical evaluation. Section VII discusses related work. Finally Section VIII provides final remarks.

## II. RUNNING EXAMPLE

In this section, we present a motivating example consisting of a simple Android app, *accentCheck*, for recording and playing raw audio data in Android. We downloaded this simple app from git (https://github.com/khan08/accentFix).

Listing 1. A code-snippet of *accentCheck*

```java
...
public class AudioRecordActivity
        extends AppCompatActivity {
    ...
    public void onClick(View v) {
        switch (v.getId()) {
            case R.id.btnStart: {
                enableButtons(true);
                startRecording();
                break;
            }
            case R.id.btnStop: {
                enableButtons(false);
                stopRecording();
                break;
            }
        }
        ...
    }
    ...
    private void startRecording() {
        ...
        recorder = new AudioRecord(...);
        recorder.startRecording();
        ...
    }
    ...
    private void stopRecording() {
        ...
        recorder.release();
        ...
    }
    ...
}
```

On startup, the app displays a graphical user interface with two buttons, the START button, which can be used to start the audio recording, and the STOP button, which can be used to stop the audio recording and replaying the audio recorded so far. Executing these operations requires acquiring and releasing the microphone using the Android APIs. Listing 1 shows a code-snippet of the app.

Method `onClick(View v)` is executed when any of the buttons is clicked. If the START button is clicked, the status of the buttons on the GUI is updated accordingly and the audio recording is initiated. Note that initiating the audio recording requires creating an object of type `AudioRecord` and invoking the method `startRecording()` on this object. If the STOP button is clicked, the status of the buttons on the GUI is also updated and the audio recording is terminated by invoking the method `release()` on the `AudioRecord`.

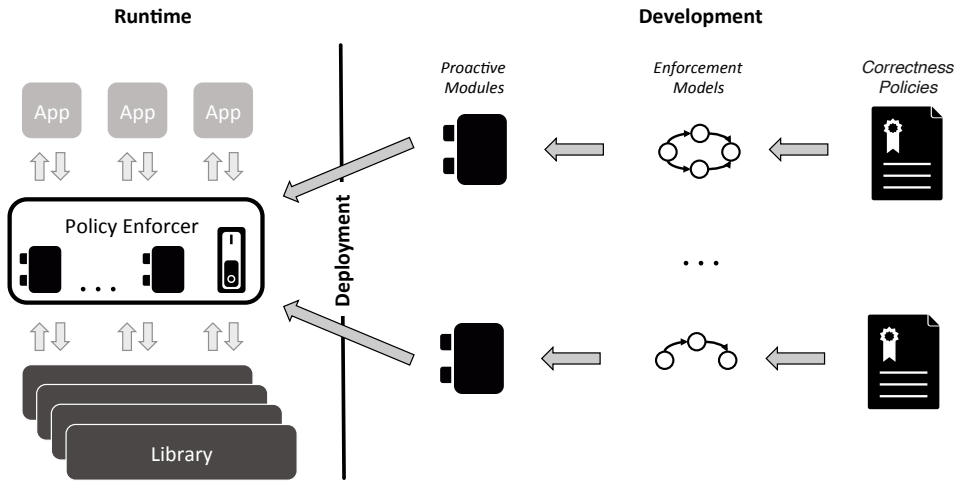

Fig. 1. Proactive Libraries.

The implementation is apparently correct. However, `stopRecording()` should also be invoked every time the `MainActivity` is no longer visible to the user. This requires implementing the `onStop()` callback method and invoking `stopRecording()` inside the method. Since this app does not implement this callback method, every time the `MainActivity` becomes invisible after the audio recording has been initiated, the audio recorder would be unaccessible from all the other apps and the app itself. On the contrary, as recommended in the Android API documentation, the `AudioRecord` should always be released when the `onStop()` method is called.

We exploit this faulty app to present the concept of proactive library and show how the proactive module associated with the `AudioRecord` API may automatically detect and heal this problem.

## III. PROACTIVE LIBRARIES

A proactive library is the ensemble of a *reactive library* and one or more *proactive module*s. The reactive library is a classic library that can be normally used as part of an application. In this paper we focus on Android libraries but the concept is generally valid for any library. The proactive modules are software components that can enforce correctness policies at runtime and that are activated by specific events, which usually correspond to the invocation of either callback methods or library methods. Invocations to *callback methods* are calls produced by the framework (in this case the Android framework) to methods implemented in the app for correctly handling lifecycle state transitions. For example, in Android there are methods that are automatically invoked by the framework when an activity is suspended, destroyed, resumed, etc [10]. Invocations to *library methods* are calls to the methods implemented by the reactive library. Correctness policies specify how a library should be used by an app relating the status of the app, traced by observing calls to callback methods, to the status of the library, traced by observing calls to library methods. Correctness policies may cover any aspect related to the usage of a library, including functional correctness, privacy, and security aspects. In this paper we focus on a specific class of correctness policies that we call *resource usage policies*. These are policies that state how the API interface of a library that controls the access to a resource should be used to prevent any misuse of the resource, which may cause misbehaviors, failures and crashes at runtime.

Note that in contrast to a regular reactive library, a proactive library may perform operations even when the library is not in use. For instance, the invocation of a callback method may trigger the execution of a proactive module, which may proactively perform some operations necessary to guarantee the correctness of the execution, regardless the direct involvement of the library in the computation. Considering the running example, a proactive module can fix the execution by forcing the release of the microphone when the `MainActivity` is no longer visible, that is, when the `onStop()` callback method is invoked, and re-assigning the microphone to the app when the activity is resumed, that is, when the `onRestart()` callback method is invoked.

Figure 1 visually illustrates how proactive libraries work, distinguishing the development and the runtime phases. At development time, developers of proactive modules start from the identification of the correctness policies. A *correctness policy* is a natural language statement that constrains the usage of the API, considering the status of the app if necessary. Any app that violates a correctness policy is faulty.

Examples of correctness policies, and more in particular resource usage policies since they refer to access to resources, for the `AudioRecord` Android library are

Resource Usage Policy 1: *"An activity that is stopped while having the control of the microphone must first release the microphone."*

Resource Usage Policy 2: *"An app cannot acquire the microphone twice."*

The first policy refers to both the status of the activity (an activity that is stopped) and the usage of the API, while the second policy only refers to the usage of the API. Both policies are valid for every app that may use the `AudioRecord` library. Note that the first policy is the one that should be exploited to heal the failures caused by the fault in the running example.

Correctness policies are then formalized and encoded as enforcement models. An *enforcement model* precisely defines how to react to correctness policies violations. We use edit automata [11] to define the enforcement models because they represent a simple finite-state based formalization that naturally supports the definition of the behavior of the proactive module in terms of the events that must be intercepted and the events that must be inserted and suppressed as a reaction to each intercepted event. The objective of the enforcement model is to *enforce* a correctness policy, even if the running apps do not satisfy it. The enforcement model is defined uniquely using the knowledge of the library API and the Android callback methods [10], which are the same for any app. Thus its definition does not require any knowledge specific to the app that uses the library. See Section IV for details about how to define enforcement models using edit automata.

An enforcement model fully describes the behavior of the corresponding proactive module, which can be thus generated from the model. Proactive modules can be deployed in any environment where the corresponding library is used. Since proactive modules are activated by the invocation of certain methods, their execution in the user environment is controlled by a *policy enforcer* that intercepts the events and dispatches them to the deployed proactive modules. The policy enforcer also controls the activation and deactivation of the proactive modules, which can be turned off and on by the user. See Section V for details about how to generate and execute a proactive module.

## IV. Enforcement Models

An enforcement model is a formal representation of the actions that must be undertaken to automatically enforce a correctness policy. In this paper, we focus on the Android ecosystem and on a specific class of correctness policies, the *resource usage policies*, which state how an API that controls the access to a resource must be used by an app. An interesting aspect about these policies is that the usage of a resource is strongly coupled with the state of the activities, that is, depending on the state of an activity there are operations that must or must not be executed.

In practice, an enforcement model traces the state of the activity by intercepting the system callbacks, which uniquely identify the state of the activity according to the Android activity lifecycle [10], traces the state of the library by intercepting the calls to the library API, and specifies the actions that must be proactively and automatically suppressed or inserted to enforce the satisfaction of a possibly violated policy. The proactive module obtained from the enforcement model guarantees the satisfaction of the policy at runtime, so that executions can be healed automatically.

In the following, we first introduce edit automata and then we discuss how edit automata can be used to specify an enforcement model.

*Edit Automata*

Edit automata have been defined by Ligatti et al. [11] to formally specify program monitors as abstract machines that can transform the sequence of operations executed by a monitored program. Edit automata are mainly used to enforce behaviors that the monitored program may fail to satisfy. Here we use edit automata to enforce correctness policies.

An edit automaton is a finite state model with the capability to modify an input sequence of actions by suppressing and inserting actions. Its behavior defines how an input sequence is incrementally changed into an output sequence. Formally it is a finite state machine whose transitions are associated with both an input action and an output sequence. The semantics of the transition is the following: when the input action is detected, the output sequence is emitted. The output sequence may be the same than the input action, which corresponds to the edit automaton not changing the execution. The output sequence may contain additional actions, which corresponds to the edit automaton inserting actions that were not part of the observed executions. The output sequence may omit the input action, which corresponds to the edit automaton suppressing the input action. Finally, the output sequence may both suppress and insert actions.

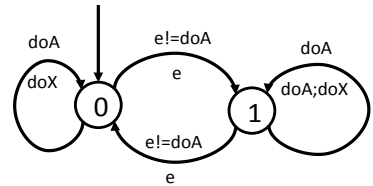

Fig. 2. A sample edit automaton.

Figure 2 shows a sample edit automaton. The symbol above a transition indicates the input symbol recognized by the transition, while the sequence below a transition indicates the output sequence emitted by the automaton when the input sequence is recognized. The input symbol can also be expressed with guard conditions, for instance `e!=doA` indicates any event different from `doA`. The arrow pointing at state 0 indicates the initial state.

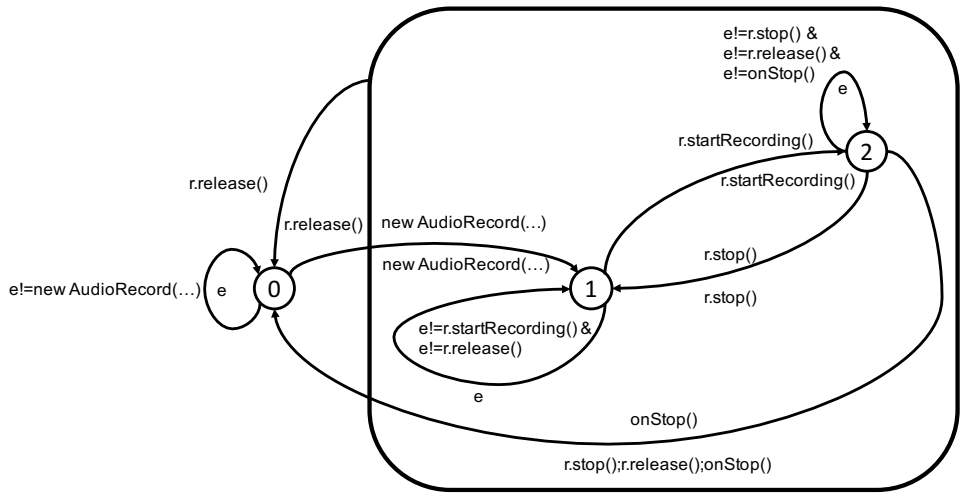

Fig. 3. Simplified enforcement model for the Resource Usage Policy 1 of the `AudioRecord`.

In the sample edit automaton, the self-loop on state 0 replaces every occurrence of the input event `doA` with the event `doX` (`doA` is suppressed and `doX` is inserted) until an event different than `doA` is observed. When an event different than `doA` is observed at state 0, the transition from state 0 to state 1 is taken and the model emits the input symbol as output (the input symbol is unmodified). At this point the self-loop on state 1 replaces every occurrence of `doA` with the sequence of events `doA` followed by `doX` (`doX` is inserted). When an event different than `doA` is observed at state 1, the transition from state 1 to state 0 is taken and the model emits the input symbol as output (the input symbol is unmodified).

For example, the edit automaton in Figure 2 transforms the sequence `doA; doA; doB; doA; doA; doC` into the sequence `doX; doX; doB; doA; doX; doA; doX; doC`.

When the edit automaton is used to specify how to enforce a correctness policy for a given library, the corresponding proactive module is a program monitor that transforms an observed sequence of actions into another sequence according to the behavior specified in the edit automaton.

*Policy Specification*

An enforcement model formally defines the actions that must be automatically undertaken to enforce policies. In the case of resource usage policies for the Android ecosystem, the enforcement models can suppress and insert actions mainly of two kinds: callback methods and API methods. Callback methods are method calls produced by the Android framework when an app changes its status. For instance, the callback method `onStop()` is automatically invoked when the running activity of the app is no longer visible, while `onDestroy()` is invoked when the activity is destroyed. Note that the set of callback methods is defined by the Android framework and does not depend on the specific app. The API methods are the methods implemented by the library associated with the proactive module. In the running example, the methods `startRecording()` and `release()` are API methods implemented by `AudioRecord`.

Figure 3 shows a slightly simplified enforcement model for the `AudioRecord` API. The prefix `r` is used to distinguish the calls to the API from callbacks. The model encodes the behavior of the proactive module for the Resource Policy 1 reported in Section III. In particular, the model forces the release of the `AudioRecord` when the activity becomes invisible without releasing the `AudioRercord`. To keep the example real but small, the enforcement model does not include the part that reassigns the `AudioRecord` to the activity once its execution is resumed. However, this part is equivalent to the one shown in Figure 3, with the difference that the action that is inserted by the model is the creation of the `AudioRecord` instead of its release and the insertion is performed when resuming the activity rather than when the activity is stopped.

In the initial state (state 0), the model waits for the creation of a new `AudioRecord`, which is the first operation performed by an app that has to access to the microphone. Every action is left unmodified by the transitions outgoing from state 0. The creation of the `AudioRecord` causes a transition to state 1. State 1 waits for the actual usage of the resource to start. That is, all the transitions leave the input action unmodified, but the observation of a call to `r.startRecording()` causes a transition to state 2. State 2 is the state that detects the violation of the resource usage policy, if any. In fact, if the app invokes `r.release()` or `r.stop()`, the model moves to states 0 and 1, respectively. Since no issue is detected, the execution is unmodified. On the contrary, if the `onStop()` callback is detected not preceded by an invocation to `release()` or `stop()`, the model modifies the execution by forcing the release of the resource emitting the output sequence `r.stop();r.release();onStop()` (see transition from state 2 to state 0).

As shown in next section, this model can be used to synthesize a monitor that enforces this behavior. Note that although the model uses universal quantifiers, the corresponding monitor does not need to intercept the execution of every method of the API and every callback. The universal quantifier must be intended as limited to the vocabulary of the edit automaton. For example, enforcing the property shown in Figure 3 only requires monitoring four API methods (`new AudioRercord()`, `startRecording()`, `release()`, and `stop()`), and one callback methods (`onStop()`).

Using edit automata to specify the behavior of a proactive module compared to directly coding the module have two important advantages: (1) it provides a compact representation of the enforcement strategy that can be more easily manipulated and modified than working at the code level, and (2) it can be composed with other finite state specifications (e.g., the specification of other policies, the app lifecycle, and the protocol of the API) to formally check its correctness. Our initial research effort has been devoted to the investigation of proactive libraries as a design solution for deploying self-healing capabilities, in particular in the Android ecosystem. Thus, we have not investigated yet the range of formal analyses offered by edit automata. Exploring this aspect, for instance to study the effect of multiple policies active at the same time, is part of our future work. For the moment we assume that the developer designs edit automata that do not interfere one with another, for example by making sure that the actions suppressed and inserted by an automaton are not monitored by the other automata.

## V. Generation of Proactive Modules

The actual proactive module that is deployed in the field and that has the responsibility to enforce a policy can be generated from the enforcement model. Since a proactive module has to respond to certain events executed by a monitored app (e.g., callbacks and calls to library methods), its implementation exploits a framework for intercepting and reacting to events. To this end, we used the Xposed framework [12], which allows to cost-efficiently intercept method invocations and change the behavior of an Android app using run-time hooking and code injection mechanisms. To avoid rooting, alternative methods, such as bytecode instrumentation [13], could be also exploited to implement the policy enforcers.

From a technical perspective, Xposed can be used to specify the Java methods that must be executed as reaction to a specific event. In our case, given an enforcement model, the events that must be hooked are the ones indicated as inputs on the transitions of the model (e.g., `new AudioRercord()`, `startRecording()`, `release()`, `stop()`, and `onStop()` for the model in Figure 3). The implementations of the methods that are executed when such events are intercepted are defined by the output emitted in each transition of the model.

An excerpt of the code corresponding to the model in Figure 3 is shown in Listing 2. The implementation is entirely contained in the `injectModule()` method that is executed by the Policy Enforcer when the proactive module is deployed. The Policy Enforcer can activate and deactivate the hooks, and thus activate and deactivate the deployed proactive modules, at any time.

In the sample code, we use two static methods provided by the Xposed framework: `XposedHelpers.findAndHookConstructor` and `XposedBridge.hookAllMethods`. They define the methods that must be executed when the invocation of a given constructor or method is intercepted, respectively. The class that implements the constructor or the method that must be intercepted is defined by the variable `classResolver`, passed as first argument (in our example these classes are `android.media.AudioRecord` and `android.app.Activity`). In case a method is intercepted, its name is passed as second argument (in our example the method is `onStop`, note that the name of the constructor does not need to be specified because it matches the name of the class). The number of arguments of these methods matches the number of arguments of the intercepted constructors or methods.

The operation that must be executed is defined by methods declared inline within the Xposed methods. In particular, the `beforHookedMethod` and `afterHookedMethod` define the operations that must be executed before and after the execution of the intercepted method. Xposed provides APIs to suppress a given method call, if necessary.

The code in the example covers two cases: the reaction to the invocation of the constructor of the `AudioRecord`, corresponding to the transition from state 0 to state 1 in Figure 3, and the reaction to the invocation of the `onStop()` callback method, corresponding to the transition from state 2 to state 0.

The method that is executed when the `AudioRecord` is created is the `afterHookedMethod` defined inline within the `findAndHookConstructor` method. The implementation checks the current state (it must be state 0 to fire the transition), emits the output (that is, the execution is unmodified), and changes the current state (to state 1). In addition the implementation encapsulates the created object into the `Manager` class. The policy enforcer automatically redirects all the invocations from the app to the resource (i.e., to the `AudioRecord`) to the `Manager`, which transparently acts as an intermediary between the app and the created object. The presence of the `Manager` is useful because the proactive module can transparently destroy and create the object stored in the `Manager`, without producing any side effect on the app.

When the `onStop` callback is invoked, the methods `beforeHookedMethod` and `afterHookedMethod` defined inline within method `hookAllMethods` are executed before and after the execution of `onStop`, respectively. Their implementation is straightforward. Method `beforeHookedMethod` enforces the release of the resource by invoking the sequence `stop()` and `release()`. The `onStop` callback is then actually executed, and finally, once `onStop` has been completed, the current state is changed,

Listing 2. A code-snippet of the Proactive Module corresponding to Resource Policy 1

```java
public class PMReleaseResource extends ProactiveModuleI {
  ...

  public void injectModule() {

    Class<?> classResolver = findClass("android.media.AudioRecord", lpparam.classLoader);

    //constructor from state 0 to 1
    XposedHelpers.findAndHookConstructor(classResolver, int.class, int.class, int.class,
                                         int.class, int.class, new XC_MethodHook() {
        ...
        @Override
        protected void afterHookedMethod(XC_MethodHook.MethodHookParam param) throws Throwable {
          ...
          if (state == 0) {
            params = param.args;
            Manager.setAudioRecord(new AudioRecord((int) param.args[0], (int) param.args[1],
                            (int) param.args[2], (int) param.args[3], (int) param.args[4]));
              state = 1;
          }
          ...
        }
    });

    //transition from state 2 to 0
    classResolver = findClass("android.app.Activity", lpparam.classLoader);

    XposedBridge.hookAllMethods(classResolver, "onStop", new XC_MethodHook() {
        @Override
        protected void beforeHookedMethod(MethodHookParam param) throws Throwable {
            if (state == 2) {
                MediaPlayerManager.release();
            }
        }

        @Override
        protected void afterHookedMethod(MethodHookParam param) throws Throwable {
            if (state == 2) {
                state = 0;
            }
        }
    });

  }

}
```

consistently with the behavior specified in the model.

## VI. EVALUATION

To evaluate proactive libraries, we considered two research questions, one about the effectiveness and one about the efficiency of the approach.

*RQ1 (effectiveness): Can proactive libraries be used to detect and heal actual library misuses in Android applications?*

With this research question, we evaluate the effectiveness of proactive libraries against actual problems reported in Android apps. As discussed in the paper, we focus on misuses of the APIs that control the access to local resources because these misuses are quite frequent [9], [4], [8], and may have a significant impact on the system, for instance affecting the other non-faulty applications that run on the same device and use the same resources.

*RQ2 (efficiency): What is the overhead introduced by proactive libraries compared to traditional reactive libraries?*

With this research question, we measure the runtime overhead that can be experienced when the proactive modules are active compared to the standard configuration that executes the reactive libraries only.

To investigate these two research questions, we selected a number of Android apps that have been reported to be affected by API misuses. We restricted the selection to open source apps to be able to investigate the faults and confirm the presence of the misuses.

To this end, we selected all the open source apps from the benchmark that Liu et al. [9] used to evaluate their static analysis technique for the detection of resource leaks. This results in three Android apps that have been reported to be affected by six resource leaks spanning five different resources, including the Bluetooth adapter, the Camera, and the Location manager. The three apps are BlueChat[2], a Bluetooth instant messaging app; fooCam[3], a camera app that can automatically take multiple successive shots with different exposure settings; and GetBack GPS[4], an application to go back to a previously visited location using GPS coordinates. We completed our benchmark with HearHere[5], an app that categorizes a tapping sound within a grid using the geometric features of the sound, which is affected by a documented resource leak on the AudioRecord[6]. We ended up with four Android apps reported to be affected by seven API misuses spanning six different resources.

TABLE I
EFFECTIVENESS OF PROACTIVE LIBRARIES

| APP | API | Resource Usage Policy | Size of EFSM Alphabet | Policy Violations |
|---|---|---|---|---|
| BlueChat | android.bluetooth.BluetoothAdapter | If enable() is invoked, invoke disable() when onDestroy() | 11 | **healed** |
| fooCam | android.hardware.Camera | If open() is invoked, invoke release() when onPause() | 29 | no violation |
| | | If startPreview() is invoked, invoke stopPreview() when onDestroy() | 29 | **healed** |
| GetBack GPS | android.location.LocationManager | If requestLocationUpdates() is invoked, invoke removeUpdates() when onPause() | 37 | **healed** |
| | android.hardware.SensorManager | If registerListener() is invoked, invoke unregisterListener() when onPause() | 18 | **healed** |
| | android.os.RemoteCallbackList | If register() is invoked, invoke unregister() when onDestroy() | 8 | no violation |
| HearHere | android.media.AudioRecord | If new AudioRecord() is invoked, invoke release() when onStop() | 13 | **healed** |

For each case, we used Appium [14], a test automation tool for mobile apps, to implement an automatic test case that can be executed to reproduce the sequence of actions that has been reported to produce the misuse. All the executions have been performed on an actual smartphone: the Samsung's Galaxy Nexus smartphone running CyanogenMod 13.0.1 OS based on the Android 6.0.1 Marshmallow mobile platform, equipped with a 1.2 GHz dual-core ARM Cortex-A9 processor, 1 GB of RAM and 16 GB of internal storage.

*RQ1 - Effectiveness*

To evaluate the effectiveness of proactive libraries, we first identified the resource usage policies relevant to the APIs involved in the reported misuses and we then defined the corresponding enforcement strategies. To identify the policies, we exploited the information about the use of the Android API reported in [9], [8], [15]. We then defined the enforcement strategy corresponding to each policy, and codified this strategy as an edit automaton, which has been finally turned into a Java proactive module.

To study the impact of the proactive libraries, we executed the test cases that should produce the misuses, and checked whether the misuses have been automatically detected and healed by the proactive modules. Table I reports the results.

[2]https://github.com/AlexKang/blue-chat
[3]https://github.com/phunehehe/fooCam
[4]https://github.com/ruleant/getback_gps
[5]https://github.com/EllingtonKirby/HearHere
[6]https://github.com/EllingtonKirby/HearHere/issues/2

Column *APP* indicates the name of the Android app used in the evaluation. Column *API* indicates the API that has been reported to be misused by the app. Note that each API maps to a different local resource.

Column *Resource Usage Policy* indicates the policy that has been reported to be violated by the app. We have written all the policies in the form "if `methodA` is invoked, invoke `methodB` when `callback`". The policy should be interpreted as: if the app invokes `methodA`, it should also invoke `methodB` when `callback` is produced by the Android framework, unless `methodB` has been already invoked before. Generally speaking, the policies require interrupting any ongoing usage of a resource if certain events occur, such as the suspension or destruction of the current activity.

Column *Size of EFSM alphabet* indicates the size of the alphabet of the edit automaton, which corresponds to the number of methods that are intercepted by the proactive module. Note that the set of methods changes based on the API and based on the specific enforcement strategy. In fact, although inserting a method invocation that has not been timely produced by an app (e.g., invoking `stopPreview()` when a destroyed activity omits to interrupt the generation of the preview) is almost the same for all the enforcement models, the way a resource that has been forcedly released can be transparently reassigned to an app when its execution is resumed changes based on the specific resource type. This aspect produces significantly different enforcement models.

Column *Policy Violations* indicates the effect of the proactive modules on the execution: *healed* indicates that the execution has been automatically corrected, while *no violation* indicates that the policy has not been violated by the execution. In none of the cases the proactive modules fail to heal an execution that violates a resource usage policy. This result confirms the effectiveness and suitability of proactive modules to enforce correctness policies.

It is worth discussing the two cases where no violation has been detected. These two cases have been reported as violating the corresponding resource usage policy in [9], where a static analysis tool is exploited to detect erroneous accesses to resources. We discovered that these two cases in reality are false positives produced by static analysis. In `fooCam`, the path that may violate the resource usage policy is infeasible, thus it can be never executed. In GetBack GPS, the path that produces the violation causes the `RemoteCallbackList` to be killed. Killing a `RemoteCallbackList` implies unregistering all registered callbacks without invoking the `unregister` method, and thus no resource leak is generated as there are no callbacks to be unregistered when the `onDestroy()` is called. Both cases are correctly represented in the enforcement models that do not report any violation when the apps are executed, contrarily to static analysis.

TABLE II
HIGHEST OVERHEAD INTRODUCED IN A SINGLE ACTION

| APP | Execution Time | | Overhead |
| --- | --- | --- | --- |
| | With Proactive Module | Without Proactive Module | |
| BlueChat | 2135 ms | 2039 ms | +4.71% |
| fooCam | 5275 ms | 5092 ms | +3.59% |
| GetBack GPS | 4013 ms | 3935 ms | +1.98% |
| HearHere | 1450 ms | 1402 ms | +3.42% |

*RQ2 - Overhead*

To measure the overhead at a fine-grained level, we measured the time taken by each individual action performed in each test case, when the proactive modules are active and when they are disabled. We repeated the execution of the test cases 50 times and computed the median execution time and overhead for each action in each test. For the majority of the actions, the use of proactive libraries does not generate any measurable overhead. This is expected because the enforcement models for the majority of the cases leave the input unaltered or perform extremely simple operations.

Only a few actions presented a negligible overhead (less than 2%), and even the actions with the highest overhead are still only marginally affected by the proactive modules. In Table II we report the action that presented the highest overhead for each application. The maximum overhead is in the order of 3-4%, with an absolute maximum equals to 4.71%. In general, the highest overhead is produced by the actions that trigger the insertion of additional actions, which usually happens when a resource must be forcedly released and when a resource must be transparently re-assigned to an app. Note that the observed overhead is almost the same time that the actual app should have taken to execute the same operation if implemented correctly, that is, the overhead introduced by forcedly releasing a resource is the same time required by the app to intentionally release that resource.

Based on these results, we can conclude that policies can be monitored efficiently. Although we did not try with a large number of policies simultaneously active, the cost of monitoring these properties is likely to be affordable. In principle, only the actions that produce the simultaneous violation of many policies might be affected by a significant overhead due to the simultaneous insertion of several actions necessary to enforce all the violated policies at once, but again the total execution time would be similar to the execution time required by a correct implementation of the app.

VII. RELATED WORK

The research described in this paper spans three distinct but related research areas: detection of API misuses, incorrect resource management, and design of self-healing systems.

Faults caused by *API misuses* are well-known to software engineers [16]. These faults have shown to be pervasive in software systems and a number of techniques to detect different kinds of misuses are available [17], [18], [19]. However, entirely preventing API misuses is still impossible, since the available techniques can only detect specific classes of misuses, such as API methods invoked in a wrong order, and cannot detect every possible misuse.

The presence of applications incorrectly interacting with an API might be exacerbated by the rapid evolution of the APIs, in fact developers often struggle adapting their software to the latest versions of the libraries. These problems are well-known to affect modern ecosystems, such as Android [20], [21]. The probability of experiencing any of these problems while using an app downloaded from a marketplace is further increased by the lack of control over the process used to develop the apps available on the marketplaces. Indeed contributors are often more focused on rapidly prototyping their apps rather than on the development of high quality applications. In a nutshell, despite the existence of techniques to assess the interaction with APIs, problems due to API misuses are still frequently experienced by users of mobile devices [4], [5], [6], [7], [8].

This paper proposes to increase the dependability of devices running apps developed by potentially untrusted third-parties by replacing traditional reactive libraries with proactive libraries, which can still be executed as reactive libraries when necessary. Proactive libraries can enforce correctness policies by proactively executing operations that influence the interaction between an API and the apps depending on it. This strategy can be effective to prevent multiple classes of problems, especially the ones related to resource usage.

Proactive libraries share the idea of extending the notion of library with ReBa [22], which is a technique for the development of libraries augmented with adapters to guarantee backward compatibility. However, ReBa and proactive libraries have different objectives and adopt different solutions. ReBa addresses problems introduced by software evolution, while proactive libraries is a general mechanism to enforce policies at runtime. Moreover ReBa is purely reactive, while proactive libraries exploit proactive modules to timely perform the actions necessary to enforce correctness. Compared to a purely reactive solution, the proactive modules can modify executions more extensively, accessing information that cannot be intercepted otherwise.

Although the concept of proactive library is general, in this paper we focused on the application of this concept to faults caused by *bad resource management*, such as apps that acquire and release resources according to wrong patterns [4], [23],

[8], [9]. Although some of these issues might be discovered with ad-hoc testing and static analysis techniques [4], [8], [9], it is generally hard to eliminate any resource management problem, covering every possible situation. Moreover, the installation of apps that use resources improperly may also impact on the apps that correctly interact with the same resources, increasing the need of solutions that can operate in the device preserving the correctness of the executions.

Proactive libraries can indeed be used to eliminate the problems that might result in the violation of a resource usage policy. Our empirical results suggest that the dependability of the interaction with the resources may significantly improve if proactive libraries are adopted to replace reactive libraries.

*Self-healing* techniques are certainly related to our work. Self-healing solutions have been studied in many different contexts, including Web applications [24], operative systems [25], and Cloud environments [26]. Some techniques also explored the idea of detecting and reacting to erroneous situations, for instance using healing connectors [27], [28] and adapters [29]. However, only few self-healing solutions have been designed to address a resource-constrained environment such as a mobile device.

Early results in the Android domain mainly concerned dynamic patch injection, automatic suppression of faulty functionalities, and healing of data loss problems. Dynamic patch injection can be used to quickly deploy fixes in Android apps [30], [31]. However, since patches are produced offline, healing is obtained with the (time-consuming) intervention of the developer. This mechanism can be useful to fix important vulnerabilities, but cannot be used for immediate and automatic healing of failing executions.

Automatic suppression mechanisms can be used to automatically detect crashes and disable the functionalities that caused a crash [4]. While this approach can be useful to prevent additional failures, it does not help fixing failing executions.

Finally, healing of data loss problems provides a strategy to prevent any loss of data due to a faulty implementation of the mechanisms to suspend and recover the execution of an app [32].

Compared to these techniques, proactive libraries provide a design solution that is complemental to mechanisms like dynamic patch injection, and potentially more general than approaches that address specific classes of failures by construction, such as data loss problems.

It is worth mentioning that monitoring and enforcement in the Android environment has been also investigated by Falcone et al. [33], but their approach does not consider the idea of designing application-independent policy enforcers that are specific to API misuses and that can be controlled by the end-users.

Finally, there exist several approaches for the automatic adaptation to a changing context of the software running in a mobile device [34], [35], [36]. Although adaptation mechanisms are usually driven by the need of optimization, and are not driven by failures, it might be possible to investigate how to apply proactive libraries to enforce adaptation policies rather than correctness policies.

## VIII. CONCLUSIONS

Program libraries implement reusable functionalities that can be conveniently integrated in many diverse applications. The correctness of this integration depends on the capability of the application to satisfy the assumptions of the library, usually consisting of constraints about when and how certain library operations can be executed. Unfortunately, satisfying these constraints is not always easy. This might be due to the complexity of the API implemented by the library, to tacit assumptions made by the library, and to the complexity of the application that uses the library. For instance, Android apps are particularly prone to faulty interactions with their libraries [9], [4], [6], [8].

To address this problem, we introduced the novel notion of proactive library, which combines a regular reactive library with multiple proactive modules that can monitor the execution and enforce the satisfaction of correctness policies about how to use the library methods at runtime. Since proactive modules are monitor that can alter executions based on correctness policies, we propose to model their behavior with edit automata, which offer a natural formalism to represent how an execution can be modified by suppressing and inserting events, which consist of method calls in our case. The edit automata can be used to obtain an implementation of the proactive modules.

We experienced our design solution for the Android environment. To investigate its effectiveness we focused on a relevant class of APIs and correctness policies, that is, APIs and policies about access to resources. The results that we obtained with multiple Android apps show that proactive libraries can efficiently enforce the specified resource usage policies.

In the future, we plan to extend our work in two main directions. Since many Android entities, such as fragments, views and services, must satisfy correctness rules similar to the ones satisfied by activities, the first direction is extending the policy enforcers with the capability to intercept the lifecycle events produced by other Android entities, thus increasing the range of properties that can be enforced. The second direction is experimenting our approach with policies of different nature, such as policies that can protect the user from security threats and privacy violations.

## ACKNOWLEDGMENT

The authors want to thank Jierui Liu, Tianyong Wu, Jun Yan, and Jian Zhang for sharing with us information about the experimental subjects used to evaluate RelFix and for answering our questions.

This work has been partially supported by the H2020 Learn project, which has been funded under the ERC Consolidator Grant 2014 program (ERC Grant Agreement n. 646867) and the GAUSS national research project, which has been funded by the MIUR under the PRIN 2015 program (Contract 2015KWREMX).

## REFERENCES

[1] S. Hyrynsalmi, A. Suominen, T. Mäkilä, and T. Knuutila, "Mobile application ecosystems: An analysis of Android ecosystem," in *Encyclopedia of E-Commerce Development, Implementation, and Management (Volume II)*. IGI Global, 2016, ch. 100, pp. 1418–1434.

[2] D. Kong, L. Cen, and H. Jin, "AUTOREB: Automatically understanding the review-to-behavior fidelity in Android applications," in *Proceedings of the ACM SIGSAC Conference on Computer and Communications Security (CCS)*, 2015.

[3] L. Li, T. F. Bissyandé, D. Octeau, and J. Klein, "DroidRA: Taming reflection to support whole-program analysis of Android apps," in *Proceedings of the International Symposium on Software Testing and Analysis (ISSTA)*, 2016.

[4] M. T. Azim, I. Neamtiu, and L. M. Marvel, "Towards self-healing smartphone software via automated patching," in *Proceedings of the International Conference on Automated Software Engineering (ASE)*, 2014, pp. 623–628.

[5] A. Banerjee, L. K. Chong, S. Chattopadhyay, and A. Roychoudhury, "Detecting energy bugs and hotspots in mobile apps," in *Proceedings of the ACM SIGSOFT International Symposium on Foundations of Software Engineering (FSE)*, 2014.

[6] L. Wei, Y. Liu, and S.-C. Cheung, "Taming Android fragmentation: Characterizing and detecting compatibility issues for Android apps," in *Proceedings of the IEEE/ACM International Conference on Automated Software Engineering (ASE)*, 2016.

[7] Z. Shan, T. Azim, and I. Neamtiu, "Finding resume and restart errors in Android applications," in *Proceedings of the ACM SIGPLAN International Conference on Object-Oriented Programming, Systems, Languages, and Applications (OOPSLA)*, 2016.

[8] T. Wu, J. Liu, Z. Xu, C. Guo, Y. Zhang, J. Yan, and J. Zhang, "Lightweight, inter-procedural and callback-aware resource leak detection for Android apps," *IEEE Transactions on Software Engineering (TSE)*, vol. 42, no. 11, pp. 1054–1076, 2016.

[9] J. Liu, T. Wu, J. Yan, and J. Zhang, "Fixing resource leaks in Android apps with light-weight static analysis and low-overhead instrumentation," in *Proceedings of the International Symposium on Software Reliability Engineering (ISSRE)*, 2016.

[10] Android, "The Activity Lifecycle," https://developer.android.com/guide/components/activities/activity-lifecycle.html, [Online; accessed 4 January 2017].

[11] J. Ligatti, L. Bauer, and D. Walker, "Edit automata: enforcement mechanisms for run-time security policies," *International Journal of Information Security*, vol. 4, no. 1, pp. 2–16, 2005.

[12] "Xposed," http://repo.xposed.info/, [Online; accessed 4 January 2017].

[13] "Redexer," http://www.cs.umd.edu/projects/PL/redexer/, [Online; accessed 8 March 2017].

[14] "Appium," http://appium.io, [Online; accessed 4 January 2017].

[15] "Android API," https://developer.android.com/guide/index.html, [Online; accessed 4 January 2017].

[16] S. Amani, S. Nadi, H. A. Nguyen, T. N. Nguyen, and M. Mezini, "MUBench: A benchmark for api-misuse detectors," in *Proceedings of the International Conference on Mining Software Repositories (MSR)*, 2016.

[17] L. Mariani, F. Pastore, and M. Pezzè, "Dynamic analysis for diagnosing integration faults," *IEEE Transactions on Software Engineering (TSE)*, vol. 37, no. 4, pp. 486–508, 2011.

[18] A. Wasylkowski and A. Zeller, "Mining temporal specifications from object usage," in *Proceedings of the IEEE/ACM International Conference on Automated Software Engineering (ASE)*, 2009.

[19] Z. Li and Y. Zhou, "PR-Miner: Automatically extracting implicit programming rules and detecting violations in large software code," in *Proceedings of the European Software Engineering Conference held jointly with the ACM SIGSOFT International Symposium on Foundations of Software Engineering (ESEC/FSE)*, 2005.

[20] M. Linares-Vásquez, G. Bavota, C. Bernal-Cárdenas, M. D. Penta, R. Oliveto, and D. Poshyvanyk, "API change and fault proneness: A threat to the success of Android apps," in *Proceedings of the Joint Meeting of the European Software Engineering Conference and the ACM SIGSOFT Symposium on the Foundations of Software Engineering (ESEC/FSE)*, 2013.

[21] M. Egele, D. Brumley, Y. Fratantonio, and C. Kruegel, "An empirical study of cryptographic misuse in Android applications," in *Proceedings of the ACM SIGSAC Conference on Computer & Communications Security (CCS)*, 2013.

[22] D. Dig, S. Negara, V. Mohindra, and R. Johnson, "ReBA: Refactoring-aware binary adaptation of evolving libraries," in *Proceedings of the International Conference on Software Engineering (ICSE)*, 2008.

[23] D. Li and W. G. J. Halfond, "An investigation into energy-saving programming practices for Android smartphone app development," in *Proceedings of the International Workshop on Green and Sustainable Software (GREENS)*, 2014.

[24] J. a. P. Magalhães and L. M. Silva, "SHÕWA: A self-healing framework for web-based applications," *ACM Transactions on Autonomous and Adaptive Systems*, vol. 10, no. 1, pp. 4:1–4:28, 2015.

[25] S. Sidiroglou, O. Laadan, C. Perez, N. Viennot, J. Nieh, and A. D. Keromytis, "ASSURE: Automatic software self-healing using rescue points," in *Proceedings of the International Conference on Architectural Support for Programming Languages and Operating Systems (ASPLOS)*, 2009.

[26] Y. Dai, Y. Xiang, and G. Zhang, "Self-healing and hybrid diagnosis in cloud computing," in *Proceedings of the International Conference on Cloud Computing (CloudCom)*, 2009.

[27] H. Chang, L. Mariani, and M. Pezzè, "Exception handlers for healing component-based systems," *ACM Transactions on Software Engineering and Methodologies (TOSEM)*, vol. 22, no. 4, pp. 30:1–30:40, 2013.

[28] ——, "In-field healing of integration problems with cots components," in *Proceedings of the International Conference on Software Engineering (ICSE)*, 2009.

[29] G. Denaro, M. Pezzè, and D. Tosi, "Test-and-adapt: An approach for improving service interchangeability," *ACM Transactions on Software Engineering and Methodologies (TOSEM)*, vol. 22, no. 4, pp. 28:1–28:43, Oct. 2013.

[30] C. Mulliner, J. Oberheide, W. Robertson, and E. Kirda, "PatchDroid: Scalable third-party security patches for Android devices," in *Proceedings of the Annual Computer Security Applications Conference (ACSAC)*, 2013.

[31] M. Zhang and H. Yin, "Appsealer: Automatic generation of vulnerability-specific patches for preventing component hijacking attacks in Android applications," in *Proceedings of the Annual Network and Distributed System Security Symposium (NDSS)*, 2014.

[32] O. Riganelli, D. Micucci, and L. Mariani, "Healing Data Loss Problems in Android Apps," in *Proceedings of the International Workshop on Software Faults (IWSF), co-located with the International Symposium on Software Reliability Engineering (ISSRE)*, 2016.

[33] Y. Falcone, S. Currea, and M. Jaber, "Runtime verification and enforcement for Android applications with RV-Droid," in *Proceedings of the International Conference on Runtime Verification (RV)*, 2012.

[34] S. Elmalaki, L. Wanner, and M. Srivastava, "CAreDroid: Adaptation framework for Android context-aware applications," in *Proceedings of the Annual International Conference on Mobile Computing and Networking (MobiCom)*, 2015.

[35] R. Rouvoy, M. Beauvois, L. Lozano, J. Lorenzo, and F. Eliassen, "MUSIC: An autonomous platform supporting self-adaptive mobile applications," in *Proceedings of the Workshop on Mobile Middleware: Embracing the Personal Communication Device (MobMid)*, 2008.

[36] F. Mancinelli and P. Inverardi, "A resource model for adaptable applications," in *Proceedings of the International Workshop on Self-adaptation and Self-managing Systems (SEAMS)*, 2006.